\documentclass{iopart}
\usepackage{graphicx,epsf}

\def\Picbox#1#2{\parbox{#1}{\epsfxsize#1\vss\vskip-1.5ex\epsffile{#2.eps}\vskip-1.5ex}}
\def\cross{\Picbox{1.5em}{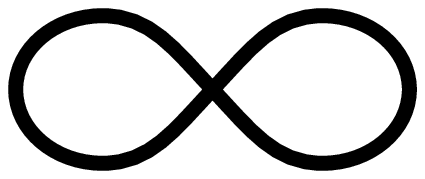}}
\def\cusp{\Picbox{1.5em}{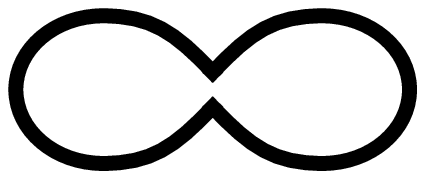}}
\def\avoid{\Picbox{1.5em}{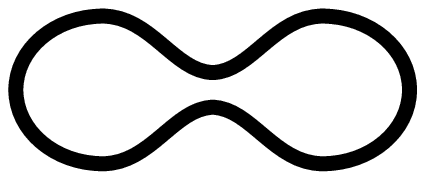}}
\def\snake{\includegraphics[width=15ex,height=1.8ex]{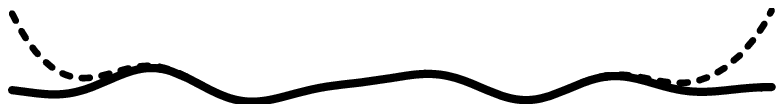}}

\def\figurename{Figure }

\begin{document}

\title{Near action-degenerate periodic-orbit bunches: A skeleton of
  chaos}

  \author{Alexander Altland$^1$, Petr Braun$^{2,6}$, Fritz
  Haake$^2$, Stefan Heusler$^3$, Gerhard Knieper$^4$ \& Sebastian
  M{\"u}ller$^5$ }

\address{\it $^1$Institut f{\"u}r Theoretische Physik,
Z{\"u}lpicher Str 77, 50937
    K{\"o}ln, Germany.  $^2$Fachbereich Physik, Universit{\"a}t
    Duisburg-Essen, 47048 Duisburg, Germany. $^3$Institut f{\"u}r Didaktik
    der Physik, Westf{\"a}lische Wilhelms-Universit{\"a}t, 48149 M{\"u}nster,
    Germany. $^4$Fakult{\"a}t f{\"u}r Mathematik, Ruhr-Universit{\"a}t Bochum,
    44780 Bochum, Germany.  $^5$School of Mathematics, University of Bristol,
    University Walk, Bristol BS8 1TW, UK.  $^6$Institute of Physics,
    Saint-Petersburg University, 198504 Saint-Petersburg, Russia.}
    \ead{Petr.Braun@uni-due.de}

\begin{abstract}
  Long periodic orbits of hyperbolic dynamics do not exist as
  independent individuals but rather come in closely packed bunches.
  Under weak resolution a bunch looks like a single orbit in
  configuration space, but close inspection reveals topological
  orbit-to-orbit differences. The construction principle of bunches
  involves close self-``encounters'' of an orbit wherein two or more
  stretches stay close. A certain duality of encounters and the
  intervening ``links'' reveals an infinite hierarchical structure of
  orbit bunches. --- The orbit-to-orbit action differences $\Delta S$
  within a bunch can be arbitrarily small. Bunches with $\Delta S$ of
  the order of Planck's constant have constructively interfering
  Feynman amplitudes for quantum observables, and this is why the
  classical bunching phenomenon could yield the semiclassical
  explanation of universal fluctuations in quantum spectra and
  transport.
  
  \vspace{0.4cm}
  
  \noindent Published in: {\it Path Integrals -
  New Trends and Perspectives: Proc  9th Int Conf (Dresden)} ed W
  Janke and A Pelster (Singapore: World Scientific) p~40 (2008)
\end{abstract}

\pacs{05.45.-a, 05.45.Mt }

 \maketitle

\section*{Introduction}
Extremely unstable motion, so sensitive to perturbation that
long-term prediction is impossible, such is the common notion of
chaos, deterministic laws {\`a} la Newton notwithstanding. As a
contrasting feature of chaos, within the continuum of unstable
trajectories straying through the accessible space, there is a
dense set of periodic orbits$^1$, and these are robust against
perturbations. Here we report that long periodic orbits associate
to hierarchically structured bunches.  --- Orbits in a bunch
are mutually close everywhere and yet topologically distinct.  The
construction principle for bunches is provided by close
self-encounters where two or more stretches of an orbit run
mutually close over a long (compared to the Lyapounov length)
distance. The self-encounter stretches are linked by orbit pieces
(``links'') of any length.  Different orbits in a bunch are hardly
distinct geometrically along the links in between the
self-encounters, but the links are differently connected in
self-encounters. Under weak resolution a bunch looks like a single
orbit, and the orbits in a bunch may have arbitrarily small action
differences. --- Orbit bunches are a new and largely unexplored
topic in classical mechanics but also spell fascination by their
strong influence on quantum phenomena: Bunches with orbit-to-orbit
action differences smaller than Planck's constant are responsible
for universal fluctuations in energy spectra$^{2-6}$, as well as
for universal features of transport through chaotic electronic
devices$^{7,8}$.

\begin{figure}[h]
\begin{center}
\includegraphics[scale=0.4]{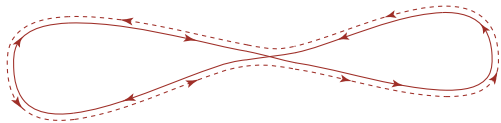}
\end{center}
\caption{{\bf Cartoon of simplest orbit bunch.}  One orbit has
  small-angle crossing which the other avoids. Difference between
  orbits grossly exaggerated.}
\label{bild1}
\end{figure}

The simplest orbit bunch, exhibited in Fig.~\ref{bild1}, is a pair of
orbits differing in an encounter of two stretches (a ``2-encounter'').
Arrows on the two orbits indicate the sense of traversal. That
elementary bunch was discovered by Sieber and Richter$^{2,3}$ who
realized that each orbit with a small-angle crossing is
``shadowed'' by one with an avoided crossing.

\begin{figure}[tbh]
\begin{center}
\includegraphics[scale=0.35]{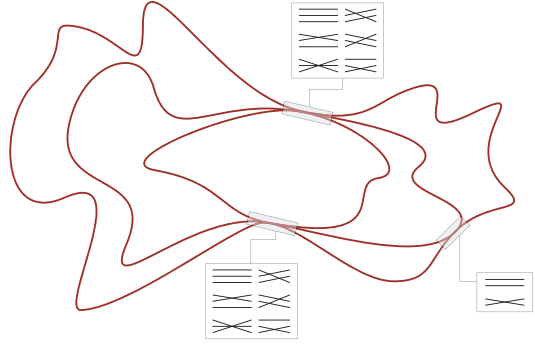}
\end{center}
\caption{{\bf Bunch of 72 orbits looking like a single orbit:} Only
blow-ups resolve intra-encounters connections distinguishing orbits;
each inter-encounter link is 72-fold, with separations yet smaller
than in encounters.}
\label{bild1a}
\end{figure}

Bunches may contain many orbits. Anticipating the
discussion below we show a multi-orbit bunch in Fig.~\ref{bild1a};
besides a 2-encounter, two 3-encounters (each involving three
stretches) are active; the various intra-encounter
connections are resolved only in the inset blow-ups.

\begin{figure}[tbp]
\begin{center}
  \includegraphics[scale=0.4,trim=0cm 0cm 0cm
  0cm]{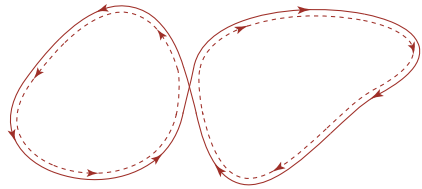}
\end{center}
\caption{{\bf Simplest pseudo-orbit,} a partner of an orbit
with a 2-encounter.}
\label{bild3}
\end{figure}

As an interesting phenomenon in bunches we meet ``pseudo-orbits'';
Fig.~\ref{bild3} depicts the prototype where the replacement of a
crossing by an avoided crossing entails decomposition of the original
orbit into two shorter orbits; the latter are then said to form a
pseudo-orbit.

\begin{figure}[h]
\begin{center}
  \includegraphics[scale=0.45,trim=0cm 0cm 0cm
  0cm]{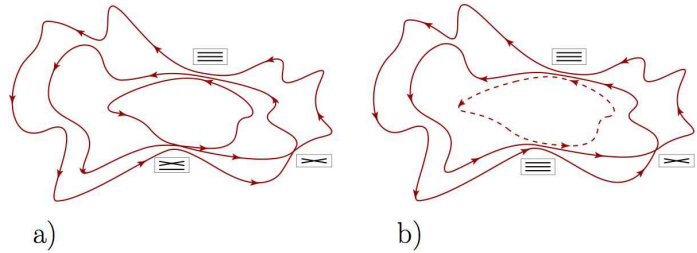}
\end{center}
\caption{{\bf Different intra-encounter connections of the bunch of Fig.~\ref{bild1a}:}
 (a) One of the 24 genuine orbits, (b) one of the 48 pseudo-orbits.}
\label{bild1b}
\end{figure}

The distinction of genuine periodic orbits and their
pseudo-orbit partners is further illustrated in Fig.~\ref{bild1b}, for
the bunch of Fig.~\ref{bild1a}.

\begin{figure}[tbph]
\begin{center}
\includegraphics[scale=0.35,angle=-90]
{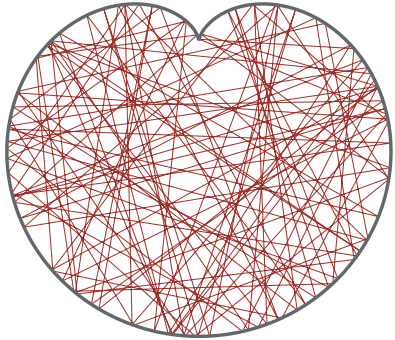}
\end{center}
\caption{{\bf Cardioid billiard} with nearly ergodic orbit.}
\label{bild2}
\end{figure}

We shall show how orbit bunches come about, illustrating our ideas
for a particle moving in a two dimensional chaotic billiard, like
the cardioid of Fig.~\ref{bild2}. The particle moves on a straight
line with constant velocity in between bounces and is specularly
reflected at each bounce.

\section*{Unstable initial value problem vs stable boundary
value problem}
 To prepare for our explanation of orbit bunches it
is well to recall some basic facts about chaos. We consider
hyperbolic dynamics where all trajectories, infinite or periodic,
are unstable (i.~e.~have non-zero Lyapounov rate $\lambda$): Tiny
changes of the initial data (coordinate and velocity) entail
exponentially growing deflections both towards the future and the
past.

Turning from the initial value problem to a boundary value problem we
may specify initial and final positions (but no velocity) and ask for
the connecting trajectory piece during a prescribed time span.  No
solution need exist, and if one exists it need not be the only one.
However, hyperbolicity forces a solution to be locally unique. Of
foremost interest are time spans long compared to the inverse of the
Lyapounov rate.  Then, slightly shifted boundary points yield a
trajectory piece approaching the original one within intervals of
duration $\sim 1/\lambda$ in the beginning and at the end, like $\snake$;
towards the ``inside'' the distance between the perturbed and the
original trajectory decays exponentially. (That fact is most easily
comprehended by arguing in reverse: only an exponentially small
transverse shift of position and velocity at some point deep inside
the original trajectory piece can, if taken as initial data, result in
but slightly shifted end points.)

When beginning and end points for the boundary value problem are
merged each solution in general has a cusp (i.~e.~different
initial and final velocities) there. If the cusp angle is close to
$\pi$ one finds, by a small shift of the common
beginning/end, a close by periodic orbit smoothing out the cusp and
otherwise hardly distinguishable from the cusped loop, like in either
 loop of Fig.~\ref{bild3}.

\section*{ Self-encounters}
 Fig.~\ref{bild2} depicts a long
periodic orbit. It appears to behave ergodically, i.~e.~it densely
fills the available space.  Fig.~\ref{bild2} also indicates that a
long orbit crosses itself many times. The smaller the crossing
angle the longer the two crossing stretches remain close; if the
closeness persists through many bounces we speak of a 2-encounter.

A general close self-encounter of an orbit has two or more stretches
close to one another. We speak of an $l$-encounter when $l$ stretches
are all mutually close throughout many bounces.  For a precise
definition one may pick one of the $l$ encounter stretches as a
reference and demand that none of the $(l-1)$ companions be further
away than some distance $d_{\rm enc}$; the latter ``encounter width''
must be chosen small compared to the billiard diameter $D$ and such
that the ensuing ``encounter length'' $L_{\rm enc}$ is much larger
than $D$.  To sum up, a close encounter is characterized by the
following order of the various length scales,
\begin{equation}\label{scales}
d_{\rm enc}\ll D\ll L_{\rm enc}\ll L\,,
\end{equation}
with $L$ the orbit length.

\section*{Orbit bunches}
 In the setting of Fig.~1 we can
formulate an important insight into chaotic dynamics$^{2,3}$. The
equations of motion allowing for the self-crossing orbit also
allow for a partner orbit which has the crossing replaced by a
narrowly avoided crossing, as shown by the dashed line in Fig.~1.
The existence of the partner follows from the shadowing
theorem$^{9}$. Arguing more explicitly, we invoke the exponential
stability of the boundary value problem mentioned above.  Namely,
the two loops of the orbit with a crossing may be regarded as
solutions of the boundary value problem with the beginning and end
at the point of crossing. We may slightly shift apart beginning
and end for each loop while retaining two junctions, as $\cross$
$\to$ $\cusp$, with nearly no change for those loops away from the
junctions. By tuning the shifts we can smooth out the cusps in the
junctions, $\cusp$ $\to$ $\avoid$, and thus arrive at the partner
orbit with an avoided crossing and reversed sense of traversal of
one loop.

Throughout the links outside the encounter the two orbits are
exponentially close. The length (and thus action) difference is
the smaller the narrower the encounter$^{2,3}$: it is quadratic
in the crossing angle, $\Delta L\propto \epsilon^2$.

The foregoing mechanism for generating partner orbits works for
$l$-encounters with any integer $l$. Each encounter serves
as a ``switch'': Its $l$ orbit stretches allow for $l!$ different
connections of the nearly fixed links outside. For
a given orbit a single $l$-encounter may and in a certain sense does
give rise to $(l!-1)$ partner orbits. The ``certain sense'' refers to
the already mentioned fact that a partner so generated may be a
pseudo-orbit, i.~e.~decompose into two or more shorter orbits (see
Fig.~\ref{bild3}).

A long orbit has many close self-encounters, some with $l=2$, some
with $l=3$, etc. Every such encounter gives rise to partner
(pseudo-)orbits whereupon many-orbit bunches come about; the number of
orbits within a bunch acquires from each close self-encounter the
pertinent factor $l!$. Fig.~\ref{bild1a} illustrates a multi-orbit
bunch; among the $(3!)^22!=72$ constituents there are 24 genuine
periodic orbits and 48 pseudo-orbits. In the various (pseudo-)orbits
of a bunch, links are traversed in different order, and even the sense
of traversal of a link may change if time reversal invariance holds
(see Fig.~\ref{bild1}).

\section*{Hierarchies of bunches}
We would like to mention two
ramifications of the concept of orbit bunches.  First, orbit
bunches form hierarchical structures, due to the near
indistinguishability of different orbits of a bunch within links.
Every link of a bunch may thus be considered as an extremely close
encounter of the participating (pseudo-)orbits, and reconnections
therein produce new longer (pseudo-)orbits; the length (and
action) of the new (pseudo-)orbit is approximately a multiple of
that of the original one. (The Sieber-Richter pair of
Fig.~\ref{bild1} makes for a pedagogical example: Considering,
say, the two left links as stretches of an encounter we may switch
these and thereby merge the two orbits.) This ``process'' of
creating ever longer orbits by selecting a link of a bunch and
treating the orbit links therein as inter-orbit encounters to
switch stretches can be continued.  Each step produces action
differences between (pseudo-)orbits exponentially smaller than the
previous step. A sequence of steps establishes an infinite
hierarchical structure, and we may see a duality of encounters and
links as its basis. --- Second, when an orbit closely encounters
an orbit from another bunch, the encounter stretches may be
switched to merge the two orbits such that the original lengths
are approximately added; clearly, the associated bunches then also
unite.

\section*{Quantum signatures of bunches} The new perspective on
classical chaos arose as a byproduct from work on discrete energy
spectra of quantum dynamics with chaotic classical limits.  As
first discovered for atomic nuclei and later found for atomic,
molecular, and many mesoscopic dynamics, the sequence of energy
levels displays universal fluctuations on the scale of the mean
level spacing. For instance, each such spectrum reveals universal
statistical variants of repulsion of neighboring levels which
depend on no other properties of the dynamics than presence or
absence of certain symmetries, most notably time reversal
invariance; correlation functions of the level density also fall
in symmetry classes. A successful phenomenological description was
provided by the Wigner/Dyson theory of random matrices (RMT); that
theory employs averages over ensembles of Hermitian matrices
modelling Hamiltonians, rather than dealing with any specific
dynamical system.

Proving universal spectral fluctuations for individual chaotic
dynamics was recognized as a challenge in the 1980's.  Only quite
recently it has become clear that orbit bunches generated by
switching stretches of close self-encounters provide the clue,
within the framework of Gutzwiller's periodic-orbit theory. The
quantum mechanically relevant bunched orbits have, as already
mentioned, action differences of the order of Planck's constant.
Using such bunches the validity of RMT predictions for universal
spectral fluctuations of individual chaotic dynamics has been
demonstrated recently$^{2-6}$.

Concurrently with the developing understanding of spectral
fluctuations just sketched, it was realized that the role of
encounters as switches is not restricted to periodic orbits but
also arises for long entrance-to-exit trajectories between
different leads of chaotic cavities$^{7,8,10-12}$. Bunches of
trajectories connecting entrance and exit leads could be invoked
to explain universal conductance fluctuations of conductors.  Like
the semiclassical work on spectra, that explanation is a welcome
step beyond RMT inasmuch as it applies to individual conductors
rather than ensembles.

\section*{Conclusion} To conclude, bunches of periodic orbits
are a hitherto unnoticed phenomenon in classical chaos, in close
correspondence to universal quantum phenomena.  System specific
behavior in mesoscopic situations is also amenable to the new
semiclassical methods$^{10-12}$ and we may expect further
application there. A semiclassical theory of localization
phenomena stands out as a challenge. --- Classical applications of
orbit bunches comprise action correlations among periodic
orbits$^{5,13}$; others could arise in a theory of the so-called
Frobenius-Perron resonances which describe the approach of ergodic
equilibrium for sets of trajectories. Similarly, orbit bunches can
be expected to become relevant for the well known cycle
expansions$^{14}$ of classical observables (where an important
role of pseudo-orbits was first noticed long since). --- The
hierarchical structure of bunches, a beautiful phenomenon in its
own right, deserves further study.

\ack This work was supported by the DFG under the SFB/TR 12. We
thank Thomas Dittrich for fruitful discussions and Stefan Thomae
for art work.

\end{document}